# The absorption refrigerator as a thermal transformer

**F Herrmann**

*Abteilung für Didaktik der Physik, Universität Karlsruhe, Germany*

**Abstract.** The absorption refrigerator can be considered a thermal transformer, that is, a device that is analogous to the electric transformer. The analogy is based on the correspondence between the extensive quantities entropy and electric charge and the intensive variables temperature and electric potential.

## 1. Introduction

The absorption refrigerator – best known as a means for food storage in recreational vehicles – is driven by a heat source unlike the compression cooling machine which is driven by an electrical motor. We examine the absorption refrigerator not for its practical importance[1], but because it is a prototypical device as is the Carnot engine.

One way of looking at the absorption refrigeration cycle is to compare it with the vapor compression cycle (Wylen and Sonntag 1965, Holman 1969). In the latter the gas is compressed, in contrast to the absorption refrigerator in which a liquid is brought from low to high pressure, which requires much less energy. Although this description is correct, it does not illustrate one aspect which might be helpful for understanding the absorption cycle. As for the Carnot cycle, the description of the absorption refrigeration cycle becomes particularly simple when looking at the entropy balance, where we can see that the absorption refrigerator is the thermal analogue of the electric transformer.

In section 2 we will bring to mind how balances of thermal engines can be formulated by means of flow diagrams. Section 3 introduces the essential aspects of the absorption refrigerator. It will be seen that, within the machine, one entropy current is flowing from a place of high to a place of low temperature thereby "pumping" another entropy current from low to high temperature.

The paper is intended for the undergraduate level.

## 2. Flow diagrams

To define what a heat engine is, it suffices to say what the engine is doing. It does not matter how the engine is doing it. An easy way to do so is by drawing a flow diagram, Fig. 1. The flow diagram shows the currents that enter and the currents that leave the machine. In addition it shows the values of the intensive quantities corresponding to the flowing extensive quantities. In the following, we always suppose that our machines work reversibly, i.e. without entropy production. This can be achieved in principle to any degree of approximation.

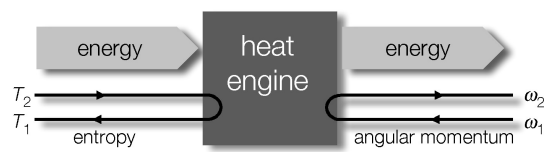

**Figure 1.** Flow diagram of a heat engine.

---

[1] The absorption refrigerator has several advantages compared to the more common compression cooling machine. It can run where no electrical energy is available, it can use the exhaust heat from thermal engines or other industrial processes, and it does not produce noise.

Let us have a closer look at Figure 1. A thermal energy flow enters the machine at the left side, and a mechanical energy flow leaves it at the right side.
Let us consider these flows separately.

1. An entropy current $I_S$ enters the machine at high temperature $T_2$ and leaves it at low temperature $T_1$. Within the machine the entropy "falls" from $T_2$ to $T_1$. Thereby a net thermal energy current

$$P_{therm} = (T_2 - T_1) I_S$$

is delivered to the machine. We say that at the entrance of the machine, the entropy is the energy carrier (Falk *et al* 1983).

2. An angular momentum current $M$ (usually called a torque) enters the machine through the mounting at the low angular velocity $\omega_1 = 0$ and leaves it at high angular velocity $\omega_2$ through the engine shaft. Within the machine, angular momentum is "lifted" from $\omega_1$ to $\omega_2$. Thereby a mechanical energy current

$$P_{mech} = (\omega_2 - \omega_1) M$$

is delivered by the machine. The energy carrier at the exit of the machine is angular momentum.

Since the energy inflow must equal the outflow, we have

$$(T_2 - T_1) I_S = (\omega_2 - \omega_1) M \tag{1}$$

Instead of considering a pure thermal engine we can use the same argument for a thermal power plant, i.e. a thermal engine plus an electric generator. Figure 2 shows the corresponding flow diagram. Instead of Eq. (1) we now have

$$(T_2 - T_1) I_S = (\varphi_2 - \varphi_1) I$$

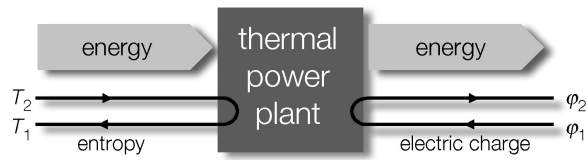

**Figure 2.** Flow diagram of a thermal power plant.

where $I$ is the electric current and $\varphi$ the electric potential.

Let us summarize in words what the flow diagram describes pictorially: An entropy current enters the power plant at a high temperature (at high "thermal potential") and leaves it at a low temperature (at low "thermal potential"). By doing so it raises electric charge from low to a high electric potential (or angular momentum from a low to a high angular velocity).

There is a similarity between a heat engine and a water wheel. Whereas within the heat engine entropy falls from high to low temperature in order to drive something ("to do work"), in a water wheel water goes from a greater to a smaller height or, expressed in more physical terms: Mass goes from a higher to a lower gravitational potential. The foregoing description of a heat engine, as well as the comparison with a water wheel, is due to Carnot (1953).

So far nothing has been said about the realization of the thermal engine. One realization, which is possible in principle but impractical has been proposed and discussed by Carnot: a cyclic process consisting of four steps, which today is called the Carnot cycle. The corresponding machine is called the Carnot engine.

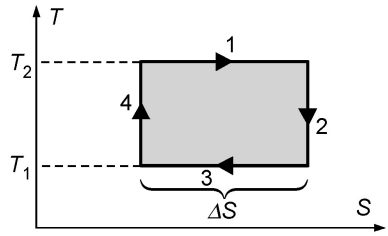

**Figure 3.** *T-S* diagram of the Carnot cycle.

The Carnot cycle is most easily characterized by its *S-T* diagram, Fig. 3. According to the direction of the cyclic process, the engine works as a thermal motor or as a heat pump. Here, we describe it as a motor. In the first step, an amount of entropy $\Delta S$ enters the engine at high temperature $T_2$. In the second step, the temperature is lowered from $T_2$ to $T_1$, while the entropy remains constant. In the third step, the engine releases the entropy $\Delta S$ while the temperature remains constant. In the forth step, the temperature increases to $T_2$ while the entropy is held constant.

The net amount of energy that is supplied thermally to the machine can be read from the *T-S* diagram. It is given by the area enclosed within the rectangle.

## 3. The absorption refrigerator as a thermal transformer

The absorption refrigerator consists of four interconnected heat exchangers, as shown in Fig. 4: the condenser, the evaporator, the absorber and the generator. The working fluid is ammonia. The ammonia flows through the four heat exchangers in the following order: condenser → evaporator → absorber → generator → condenser →… The condenser and the absorber are at ambient temperature $T_{amb}$. The evaporator is at low temperature $T_{low}$, the generator at high temperature $T_{high}$.

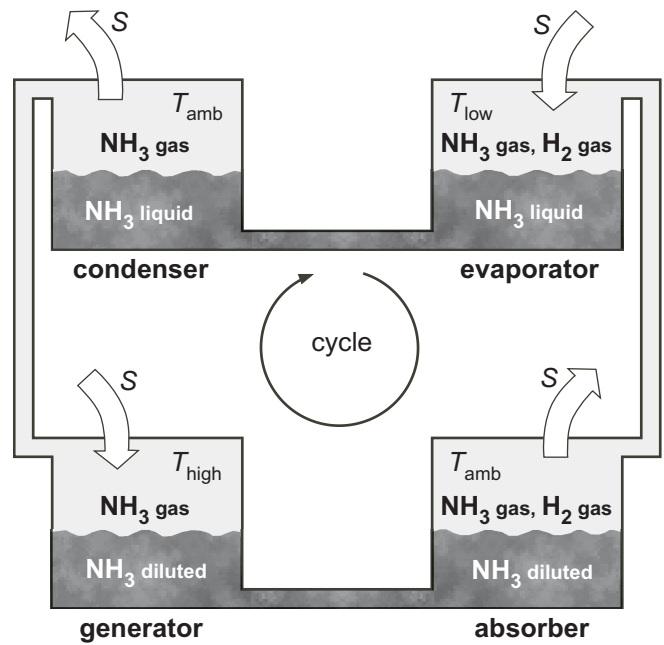

**Figure 4.** Schematic arrangement of the absorption refrigerator.

The ammonia exists in two different phases inside each of the four heat exchangers. Liquid and gaseous ammonia coexist in the condenser and the evaporator. There is gaseous ammonia and ammonia in hydrous solution in the absorber and the generator.

In addition to the ammonia, the gaseous phase in the evaporator and in the absorber also contains hydrogen. The hydrogen does not participate in the cyclic process. It ensures only that the total pressure in the evaporator is the same as in the condenser.

The transport of ammonia from the absorber to the generator could in principle proceed via diffusion. Because diffusive processes are slow, the ammonia is more effectively transferred by means of the so-called absorption circuit. The liquid ammonia-water mixture circulates in such a way that the ammonia proceeds convectively from the absorber to the generator. This liquid current is driven either by a pump or by natural convection. In the present context the absorption circuit is no more than a technical trick that accelerates the flow of the working fluid. Because we are interested only in understanding the basic principles of the machine, we omit the absorption circuit and assume that the ammonia transport from the absorber to the generator occurs by diffusion.

We are now prepared to understand the principle of the absorption machine. We follow a portion of ammonia on its way through the four stations of the machine. It is essential to consider the entropy balance: Where, and at what temperature does the portion pick up entropy, and where does it release entropy? We consider the process in a *T-S* diagram, Fig. 5.

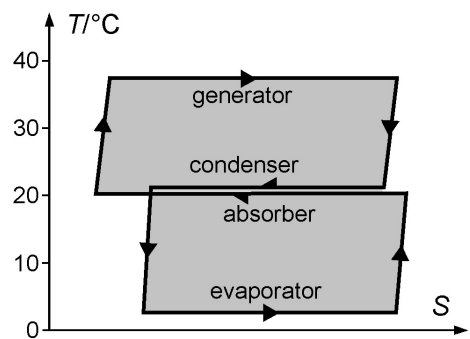

**Figure 5.** *T-S* diagram of the absorption refrigerator.

We begin with the gaseous phase of the condenser. Within the condenser the gaseous ammonia portion changes into the liquid phase. Because the entropy content of a given amount of the liquid is much lower than in the gaseous phase, the ammonia has to release entropy. This entropy leaves the machine at ambient temperature. The liquid ammonia portion next proceeds to the evaporator.

Because the evaporator is at lower temperature than the condenser, the portion of ammonia has to release entropy on its way from the condenser to the evaporator. We suppose that it does so via the walls of the pipe that connects both heat exchangers. This amount of entropy is small compared with the entropy exchanged in any of the four heat exchangers and can be neglected for our purposes.

In the evaporator the ammonia portion changes into the gaseous phase again. For that purpose it has to absorb entropy at low temperature $T_{low}$.

From the evaporator the ammonia portion proceeds to the absorber where it dissolves in the ammonia-water solution, thereby releasing entropy at ambient temperature.

From the absorber, the ammonia portion proceeds to the generator where it transforms from the solute into the gaseous phase, thereby absorbing entropy at $T_{high}$.

After this somewhat laborious step-by-step description, let us consider the entire cycle in the *T-S* diagram in the whole, Fig. 5. We no longer ask what the ammonia portion is doing, we only ask for the entropy exchanges. Because the entire cyclic process is a steady state process, the diagram of Fig. 5 also tells us how much entropy is exchanged in each of the four exchangers in a given time interval. It is seen that within the machine a certain amount of entropy proceeds from the high temperature $T_{high}$ to the intermediate temperature $T_{amb}$ whereby another entropy portion is "pumped" from the low temperature $T_{low}$ to the intermediate temperature $T_{amb}$.

The *T-S* diagram also displays the energy balance. The cycle consists of two loops. As the ammonia portion makes a complete cycle, it passes clockwise through one of these loops and counterclockwise through the other. The area of the upper loop corresponds to the energy absorbed by the ammonia portion; the lower loop corresponds to the released energy. Because the total amount of energy exchanged by the refrigerator in a given time interval is zero, the absolute values of the areas of the two loops must be equal. If we take into account that the process runs through the loops in opposite directions and accordingly attribute opposite signs to the areas of the two loops, we also can say that the total area is zero.

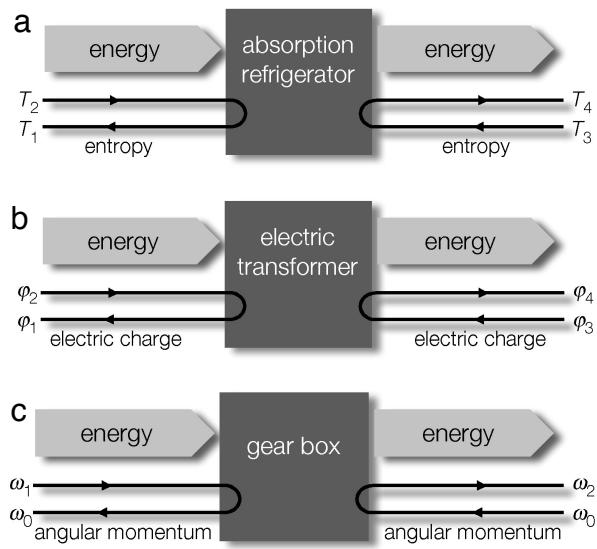

**Figure 6.** Flow diagrams of (a) absorption refrigerator, (b) electric transformer and (c) gear box.

Figure 6a shows the flow diagram of the absorption refrigerator. When comparing it with the flow diagram of the electric transformer, figure 6b, we see that these devices display a great similarity (Fuchs 1969). (The fact that an electric transformer only works with alternating electric currents is not important in this context.)

Let us remind that the mechanical analogue of the electric transformer (and the absorption refrigerator) is the gear box. An angular momentum currents enters it via the input shaft. In the gear box, this current flows from the high angular velocity of the shaft to the angular velocity 0 of the mounting. Simultaneously, another angular momentum current is "pumped" from a lower to a higher angular velocity. This is the current that leaves the gear box via the output shaft. Figure 6c shows the corresponding flow diagram.

## References

Carnot S 1953 *Réflexions sur la puissance motrice du feu et sur les machines propres à développer cette puissance, Nouvelle èdition* (Paris: Librairie scientifique et technique, A. Blanchard) p 28: "D'après les notions établies jusqu'á présent, on peut comparer avec assez de justesse la puissance motrice de la chaleur à celle d'une chute d'eau….La puissance motrice d'une chute d'eau dépend de sa hauteur et de la quantité du liquide; la puissance motrice de la chaleur dépend aussi de la quantité de calorique employée, et de ce qu'on pourrait nommer, de ce que nous appellerons en effet la hauteur de sa chute, c'est-à-dire de la différence de température des corps entre lesquels se fait l'échange du calorique." ("According to the concepts presently established, one can compare with much correctness the motive power of heat with that of a waterfall... The motive power of a waterfall depends on its height and on the amount of the liquid; the motive power of the heat also depends on the amount of caloric that is employed, and of what on could call, and what we indeed shall call the height of its fall, i.e. the temperature of the bodies between which the exchange of the caloric is done.") [Translation by F. H.]

Falk G, Herrmann F and Schmid G Bruno 1983 Energy forms or energy carriers? *Am. J. Phys.* **51** 1074-7
Fuchs H 1996 *The Dynamics of Heat* (New York: Springer) p 553-4
Holman J P 1969 *Thermodynamics* (New York: McGraw-Hill) p 361
Van Wylen G J and Sonntag R E 1965 *Fundamentals of Classical Thermodynamics* (New York: Wiley) p 288